# Impact of $BaTiO_3$ nanonoparticles on pretransitional effects in liquid crystalline dodecylcyanobiphenyl

[1,2]S. J. Rzoska (*), [2]S, Starzonek, [2]A. Drozd-Rzoska, [3]K. Czupryński,

[1]K. Chmiel, [1]G. Gaura, [1]A. Michulec, [1]B. Szczypek, [1]W. Walas.

[1]Silesian Intercollegiate Centre for Education and Interdisciplinary Research & Institute of Physics, University of Silesia, ulica 75. Pułku Piechoty 1A, 41-500, Chorzów, Poland

[2]Institute of High Pressure Physics, Polish Academy of Sciences, ullica Sokołowska 29/37, 01-142 Warsaw, Poland

[3]Military University of Technology, Institute of Chemistry, ulica Kaliskiego 2, 00-908 Warsaw, Poland

(*) Corresponding author: sylwester.rzoska@gmail.com

## ABSTRACT

The pretransitional behavior of dodecylcyanobiphenyl (12CB, *isotropic - smectic A – solid* mesomorphism) with $d$=50nm $BaTiO_3$ nanoparticles (NPs) linked to the cubic phase was monitored via temperature studies of dielectric constant. Tests were carried out in the isotropic, liquid crystal mesomorphic, and solid phases. For each phase transition the same value of the "critical" exponent $\alpha \sim 0.5$ was obtained, including nanocolloids. All phase transitions show the weakly discontinuous nature. The temperature metric of the discontinuity $\Delta T$ notably decreases when adding nanoparticles. The addition of nanoparticles first decreases the dielectric constant by approximately 50 % in comparison with pure 12CB, but already for a concentration ~ x = 0.4 % NP an increase over 50 % takes place. It is notable that for the latter concentration unique hallmarks of the pretransitional effect emerge also for the solid – mesophase transition. All these indicate the important impact of nanoparticles on multimolecular, mesoscale fluctuations.

## I. INTRODUCTION

Nanocolloids are one of the most advanced materials with properties important for a variety of innovative applications. They are obtained by adding to a liquid even a very small amount of solid nanoparticles (NPs) [1-4]. Liquid crystals (LC) have attracted particular attention as a dispersing medium for nanoparticles due to the richness of phase transitions, the importance of mesoscale structures, and enormous sensitivity to exogenic impacts, such as pressure or electric field [5-10]. Consequently, one can expect that the addition of nanoparticles can notably tailor properties of liquid crystals without new chemical synthesis. Generally, it is expected that nanoparticles, with a length scale below *100 nm*, can act as a specific additive moderator of molecular properties of the liquid crystalline host. It is indicated that the electric field associated with NPs can influence the local orientational ordering, increase the isotropic-nematic (I-N) transition temperature or decrease the switching voltage [1, 2, 10-12]. All these have led to an increase of theoretical and experimental interest in LC-based nanocolloids [17-29]. Regarding experiments, particularly important are dielectric studies since they are strongly associated with practical implementations and their results can be compared with fundamental models of the physics of liquid crystal [10-30].

Surprisingly, the evidence related to basic pretransitional properties in LC-NP nanocolloid is still lacking. These include such fundamental properties as "critical" exponents, the discontinuity of the isotropic – mesophase transition or the range of temperatures dominated by pretransitional phenomena. Such a reference is fundamental for the ultimate modeling of such composite systems.

The aim of this work is a preliminary study focusing on this enigma. It is based on dielectric permittivity investigations in dodecylcyanobiphenyl (12CB) with the addition of nanoparticles ($BaTiO_3$, $d \approx 50nm$). For 12CB the $Isotropic \longleftrightarrow SmecticA \longleftrightarrow Solid$ phase sequence takes place [5, 6]. This Rapid Communication not only focuses on the isotropic to mesophase transitions, but also shows evidence of a dielectric permittivity anomaly for the $SmA \to I$ pretransitional effect. Finally, the emergence of a $Solid \to SmA$ pretransitional effect in the LC-NP nanocolloids was found.

**Basics for the isotropic – mesophase transition**

Rod-like LC materials constitute one of the most important materials for modern civilization due their enormous significance for practical implementations matched with fundamental properties that can be relatively simply modeled [5]. In these materials a unique sequence of phase transition associated with freezing and melting of subsequent elements of symmetry takes place. This mesomorphism is mostly associated with weakly discontinuous phase transitions. The simplest example is the orientational freezing related to the isotropic liquid – nematic (*I-N*) transition or the isotropic – smectic A (*I-SmA*) transition where additionally the one-dimensional positional ordering freezes [5]. Regarding fundamentals, the unique position has the pretransitional effect for the I-N phase transition, whose description led to the formulation of the Landau – de Gennes (LdG) model, which is one of the most important theoretical concepts for the soft matter category. Basic studies in this domain were related to the Cotton-Mouton effect, the Kerr effect, light scattering, and later the nonlinear dielectric effect [5-9, 30] measurements in the isotropic phase on approaching the isotropic – nematic transition. Experimental characteristics of the mentioned properties are directly coupled to premesomorphic fluctuation and show strong pretransitional anomalies. Following these studies, the most important metrics of the I-N transition are considered the values of the power exponent characterizing the pretransitional effect or the temperature measure of the discontinuity ($\Delta T^*$) [1-7, 30].

Bradshow and Raynes [31] found a notable pretransitional effect also for the static dielectric permittivity (dielectric constant) in an *n*-cyanobiphenyl (*n*CB from $n = 5$ to $n = 12$) homologous series, where both I-N and I-SmA transitions occur. The dielectric constant is one of the most important characteristics of LC materials, since it is essential for describing interactions with the electric field and thus is important for display technology. In the nematic phase measurements of the dielectric constant constitute a basic tool for determining the behavior of the order parameter. The successful parametrization of the pretransitional anomaly for *I-N* and *I-SmA* transitions is possible [32-36]:

$$\varepsilon(T) = \varepsilon^* + a^*\left(T - T^*\right) + A^*\left(T - T^*\right)^{\phi} \quad , \qquad T > T^C \qquad (1)$$

where $\left(\varepsilon^*, T^*\right)$ are the loci of the hypothetical continuous phase transition, $T > T^C = T^* + \Delta T^*$, $\Delta T^*$ is the temperature metric of the discontinuity of the isotropic to mesophase transition and $T^C$ is for the isotropic – mesophase transition clearing temperature. The power exponent $\phi = 1 - \alpha$, where $\alpha$ is related to the specific heat critical anomaly. Here $A^*$ and $a^*$ are constant amplitudes.

Experiments for I-N, I-N*, I-SmA and I-SmE phase transitions showed that for this case the exponent $\phi \approx 0.5$ and then $\alpha \approx 0.5$. The theoretical derivation of Eq. (1) from the extended LdG model is given in Ref. [7]. It is notable that the basic LdG approach is only for the I-N transition and yields no anomaly for the dielectric constant in the isotropic phase.

The evidence for the pretransitional effect of dielectric properties for a similar mesophase to isotropic transition is poor. It covers only the case of the $N \rightarrow I$ transition where the pretransitional anomaly was detected for the behavior of the diameter, namely [35] :

$$\varepsilon_{diam.}(T) = \frac{1}{3}\varepsilon^{||} + \frac{2}{3}\varepsilon^{\perp} = \varepsilon^{**} + a_{diam.}\left(T - T^{**}\right) + A_{diam.}\left(T - T^{**}\right)^{\phi}, \quad T < T^C \qquad (2)$$

where $\varepsilon^{\parallel}$ and $\varepsilon^{\perp}$ are dielectric constants for perfectly ordered nematics in the directions parallel and perpendicular to the long axis of the rod-like molecule, respectively $A_{diam.}$ and $a_{diam.}$ are constant amplitudes, $T^{**} = T^C + \Delta T^{**}$ is the extrapolated locus of the hypothetical continuous nematic to isotropic phase transition and $\Delta T^{**}$ is the temperature metric of the discontinuity for this transition.

In the tested case in Ref. [35] of *n*-octyloxycynobiphenyl (8OCB, the permanent dipole moment parallel to the long axis) the exponent $\phi = 1 - \alpha \approx 0.5$. The relation (2) may be considered as the parallel of Eq. (1) for the nematic phase. Its form also resembles the description of a diameter of coexistence curves in binary mixtures of limited miscibility. So far there is no evidence for the $SmA \to I$ pretransitional effect of the dielectric constant, despite the fact that such studies do not involve a very strong magnetic field, which is necessary in the case of the nematic phase [5, 30].

There is a growing number of reports on dielectric properties of LC-NPs colloids. This is associated with developing theoretical studies, including such basic approaches for the physics of liquid crystals as the Maier-Saupe or Landau-de-Gennes models [10-29]. However, there are neither theoretical predictions nor experimental results regarding the form of pretransitional effects.

## II. EXPERIMENT

The tested sample of 12CB was synthesized and deeply purified to reach the minimal electric conductivity at Military University of Technology in Warsaw, Poland. Dodecylcyanobiphenyl exhibits the Isotropic-(330.8 K)-Sm-A (307.3 K)-solid mesomorphism. Dodecylcyanobiphenyl belongs to the one of the most "classical" liquid crystalline homologous series *n*CB, regarding both fundamentals and practical implementations. For all *n*CBs molecules have approximately a rod-like form, with a permanent dipole moment parallel to the long axis. Its approximate value is equal to $\mu \approx 5D$. All these characteristics lead to a large anisotropy of dielectric permittivity for the perfectly ordered sample, namely $\varepsilon_{\perp} \approx 4$ and $\varepsilon_{||} \approx 19$. $BaTiO_3$ nanoparticles (*d* = 50nm) were purchased from Research Nanomaterials (USA). All $BaTiO_3$ phases exhibit ferroelectricity except the cubic phase, which is applied in our research of nanoparticles. Mixtures of 12CB and $BaTiO_3$ NPs were sonificated with ultrasound frequency $f$ = 42 kHz for a few hours in the isotropic phase (60 $^{o}$C) until homogenous mixture was obtained for study. No sedimentation for at least 24 h was observed, hence the tested nanocolloid did not contain additional stabilizing agents. The impedance analyzer (Solartron SI 1260) enabled high resolution determination of the dielectric constant with permanent five digit resolution. For the static domain between 1 kHz and 1 MHz, changes of the real part of the dielectric permittivity $\varepsilon'(f)$ were below 1%, so measurement of the dielectric constant $\varepsilon = \varepsilon'(f)$ was carried out for $f$ = 50 kHz. The temperature was controlled by a Julabo thermostat (with external circulation, 20 L volume), The temperature stability of samples was better than 0.02 K. Samples were placed in the measurement capacitor made from Invar, with a $d$ = 0.2 mm gap and diameter 2$r$ = 20 mm. Its design is given in Ref. [36]. The quartz ring was used as the spacer. This enabled observation of the interior of the capacitor. The latter and the macroscopic gap of the capacitor made it possible to avoid bubbles distorting the results. For each concentration of nanoparticles at least three series of measurements were carried out. Additionally, properties of samples were controlled after measurements (phase transition temperatures and sedimentation).

## III. RESULTS AND DISCUSSION

Figure 1 shows results of dielectric constant measurements in the isotropic phase, LC mesophase and solid phases of 12CB: pure ($x$ = 0) 12CB and for nanocolloids with the additions of $x = 0.1\%$, $0.2\%$, $0.3\%$ of and 0.4% (mass weigh fraction) of $BaTiO_3$ nanoparticles. Higher concentrations were not tested due to emerging sedimentation hallmarks. The addition of NPs had a negligible impact on the clearing temperature (isotropic – mesophase transition) as visible in Fig. 1 and in Table I. This behavior is contrary to existing model expectations and experimental evidence, available mostly for pentylcyanobiphenyl (5CB, I-N transition) and E7 (mixture of four LC compounds, including 5CB and 8OCB, with an I-N transition)-based nanocollids, where the notable rise of the clearing temperature was detected. However, in Fig. 1 a notable influence of nanoparticles on the mesophase – solid (M-S) phase transition temperature is visible.

At first glance in Fig. 1, the pretransitional effect in the isotropic phase is weak. However, this is only impression associated with the applied scale. The focused insight in Fig. 2 reveals a notable pretransitional anomaly in the isotropic phase, associated with the crossover from the domain associated with the prevalence of chaotically oriented permanent dipole moments ( $d\varepsilon'/dT < 0$ ) to the domain dominated by antiparallel ordering of permanent dipole moments ( $d\varepsilon'/dT > 0$ ) on approaching the clearing temperature. The latter is related to the increasing impact of a rod-like molecules with a premesomorphic arrangement within pretransitional fluctuations. The latter is also associated with antiparallel ordering of permanent dipole moments, caused by the statistical equivalence of $\vec{n}$ and $-\vec{n}$ directors for orientational (nematic) ordering [30]. The loci of the crossover, i.e., $d\varepsilon'/dT = 0$, strongly depends on the concentration of nanoparticles (Fig. 2). However, in each case the pretransitional anomaly is well described by Eq. (1), with the same value of the power exponent $\phi = 1 - \alpha \approx 0.5$. Notable is the strong dependence of the discontinuity of the isotropic to mesophase transition ( $\Delta T^*$ ) on the concentration of nanoparticles. Results of fitting are shown by solid, red curves in Figs. 1 and 2. Related parameters are collected in Table I. To describe the pretransitional behavior on approaching the clearing temperature within the LC mesophase the counterpart of Eq. (1) was used, namely

$$\varepsilon(T) = \varepsilon^{**} + a^{**}\left(T^{**} - T\right) + A^{**}\left(T^{**} - T\right)^{\phi} \quad , \quad T < T^{C} \qquad (3)$$

where $\left(\varepsilon^{**},T^{**}\right)$ are the coordinates of the hypothetical continuous phase transition, $T^{**}=T^{C}+\Delta T^{**}$, $\Delta T^{**}$ is the temperature metric of the discontinuity of mesophase to isotropic transition and $A^{**}$ and $a^{**}$ are constant amplitudes.

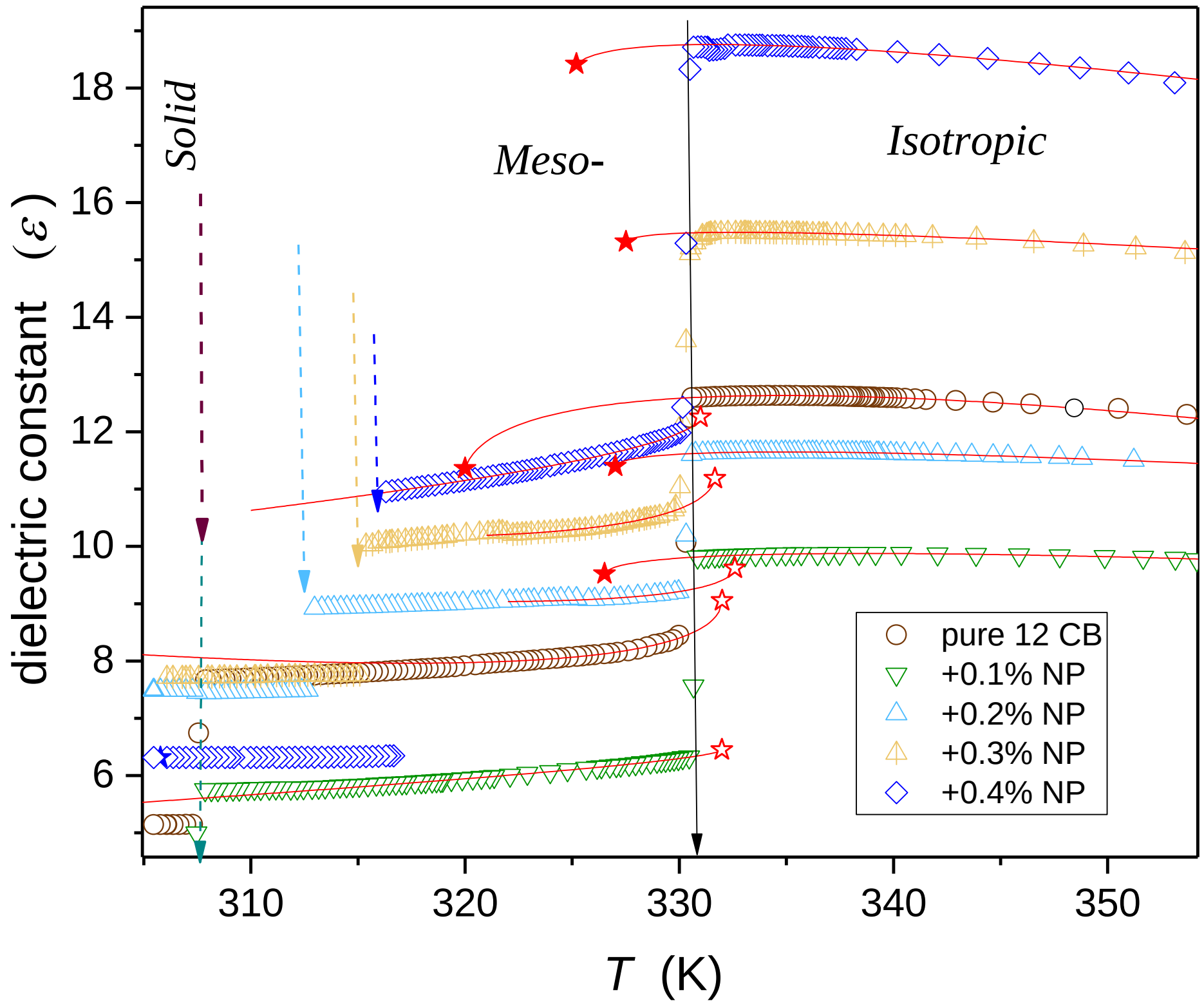

**FIG. 1. (Color online)** Temperature changes of the dielectric constant in 12CB and its colloids with $BaTiO_3$ nanoparticles. Their concentrations are given in percent mass fraction. The solid arrows indicate the isotropic-mesophase transitions (clearing temperatures $T_C$), which are approximately the same for all concentrations. Dashed arrows are for mesophase-solid transitions, related to different concentrations of NPs. Red curves show the experimental data according to Eqs. ( 1) and (3) , with the parameters collected in Table I. Closed stars are for the extrapolated temperature of the hypothetical isotropic to mesophase continuous phase transition, and open stars are the same for the mesophase to isotropic transition.

The results of such a description are shown in Fig 1 (red curves) and the parameters are collected in Table I. Also, in this case the description with the power exponent $\alpha = 0.5$ is possible. It is worth indicating that the discontinuities $\Delta T^{**}$ are notably smaller than $\Delta T^{**}$.

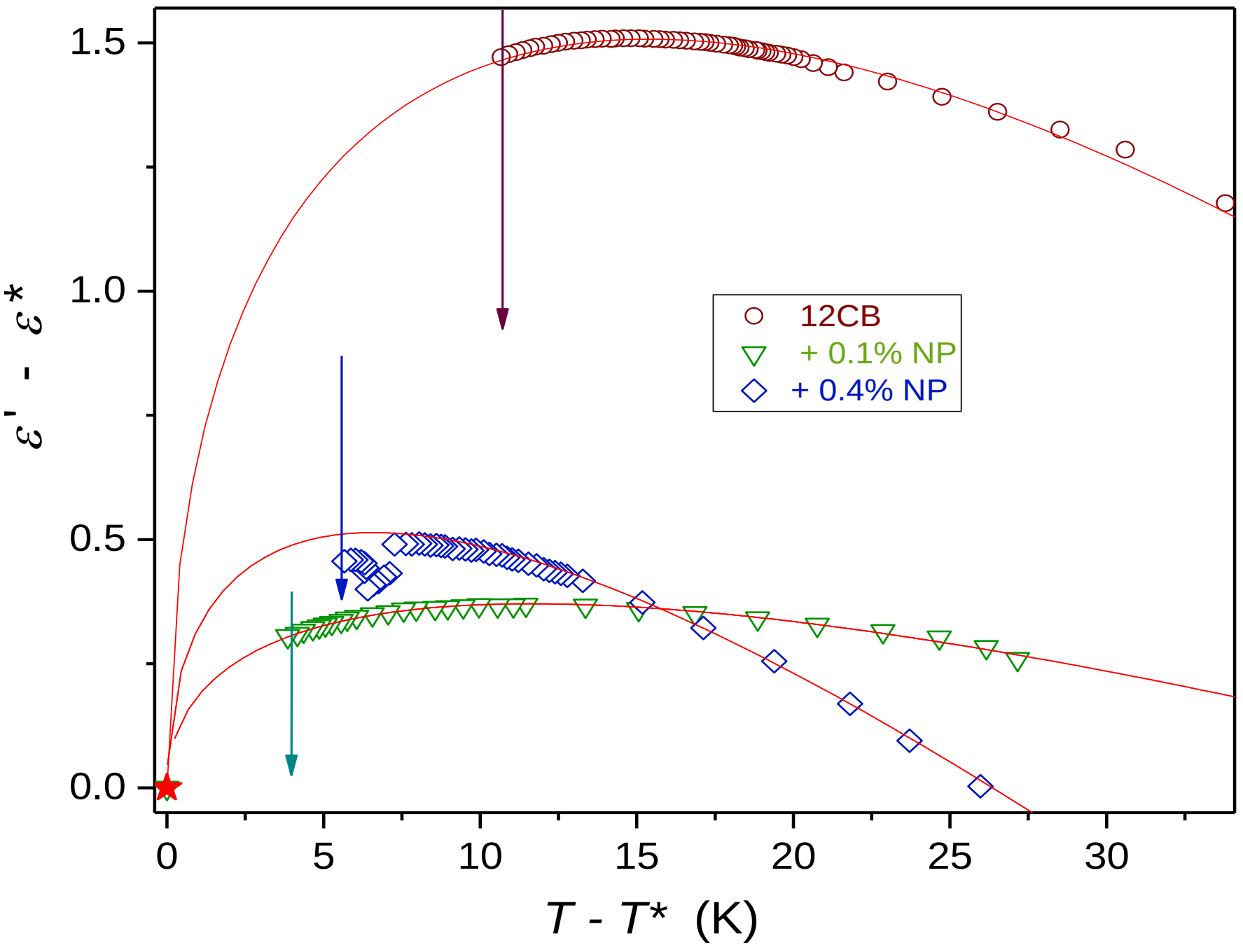

**FIG. 2. (Color online)** Temperature changes of the dielectric constant in the isotropic liquid state of 12CB and its colloids with $BaTiO_3$ nanoparticles in normalized scales referred to as the hypothetical continuous phase transition $(\varepsilon^*, T^*)$. Solid red curves are described by Eq. (1) with the parameters from Table I. Vertical arrows indicate clearing temperatures. The distances between vertical arrows and the red star indicate values of discontinuities of isotropic-mesophase transitions $\Delta T^*$.

When adding nanoparticles both values of both discontinuities $\Delta T^*$ and $\Delta T^{**}$ notably decrease, i.e., the phase transition approaches the continuous one.

It is notable that Eqs. (1) and (3) are associated with multiparameter fitting and related pretransitional anomalies are not particularly pronounced (Figs. 1 and 2). All these have notably worse reliability of the fitting. To solve this Gordian Knot, the supporting derivative analysis was used. The results are presented in Fig. 3. The number of fitting parameters decreases and leads to a strong manifestation of the pretransitional behavior. Following Eqs. (1) and (3), one obtains

$$\frac{d\varepsilon}{dT}(T) = a^*T + A^*(1-\alpha)(T - T^*)^{-\alpha} \tag{4}$$

for the $I \to M$ transition and

$$\frac{d\varepsilon}{dT}(T) = a^{**}T + A^{**}(1-\alpha)(T^{**} - T)^{-\alpha} \tag{5}$$

for the $M \to I$ transition.

These equations recall the link between $d\varepsilon/dT$ and the critical anomaly of the specific heat $c_p \propto (T - T_C)^{-\alpha}$, where $T_C$ is the critical temperature, first indicated by Mistura [37].

**TABLE I.** Results of fitting of dielectric constant $\varepsilon(T)$ experimental data for liquid crystalline 12CB and its nanocolloids with $BaTiO_3$ nanoparticles (Fig. 1) via Eq. (1). The input parameter were taken from fitting of $d\varepsilon(T)/dT$ via Eqs. 3, 4 (Fig. 3).

| Phase transition, $x$: (wt. % of NP), phase transition temperature $T$ (K) | $\varepsilon^*, \varepsilon^{**}$ | $T^*$, $T^{**}$ | $a^*, a^{**}$ | $A^*, A^{**}$ | $\alpha = 1 - \phi$ |
|---|---|---|---|---|---|
| $I \to LC$, $x$ = 0 %, 330.5 | $11.12_4$ | 319.9 | $-0.10_2$ | $0.78_6$ | 0.5 |
| $I \to LC$, $x$ = 0.1 %, 330.8 | $9.50_7$ | 326.5 | $-0.030_8$ | $0.21_6$ | 0.5 |
| $I \to LC$, $x$ = 0.2 %, 330.3 | $11.38_0$ | 327.0 | $-0.034_3$ | $0.19_2$ | 0.5 |
| $I \to LC$, $x$ = 0.3 %, 330.3 | 15.30 | 327.5 | $-0.034_9$ | $0.15_9$ | 0.5 |
| $I \to LC$, $x$ = 0.4 %, 330.6 | $18.25_9$ | 325.0 | $-0.077_2$ | $0.38_9$ | 0.5 |
| $LC \to I$, $x$ = 0 %, 330.5 | $9.07_2$ | 332.0 | $0.076_5$ | $-0.58_3$ | 0.5 |
| $LC \to I$, $x$ = 0.1 %, 330.8 | $6.46_6$ | 332.0 | $0.016_8$ | $-0.09_0$ | 0.5 |
| $LC \to I$, $x$ = 0.2 %, 330.3 | $9.65_3$ | 332.3 | $0.051_2$ | $-0.35_6$ | 0.5 |
| $LC \to I$, $x$ = 0.3 %, 330.3 | $11.26_7$ | 331.7 | 0.0725 | $-0.56_6$ | 0.5 |
| $LC \to I$, $x$ = 0.4 %, 330.6 | $12.28_6$ | 331.0 | $-0.015_3$ | $-0.29_2$ | 0.5 |
| $S \rightarrow LC$, $x = 0.4$ %, 316.1 | ~~~~ | 318 | ~~~~ | ~~~~ | 0.5 |

Consequently, the results exemplified in Fig. 3 were fitted and analyzed via Eqs. (4) and (5). The parameters obtained were then used as the input data for the analysis of experimental data presented in Fig. 1 via Eqs. (1) and (3). However, this final fitting was reduced solely to one parameter.

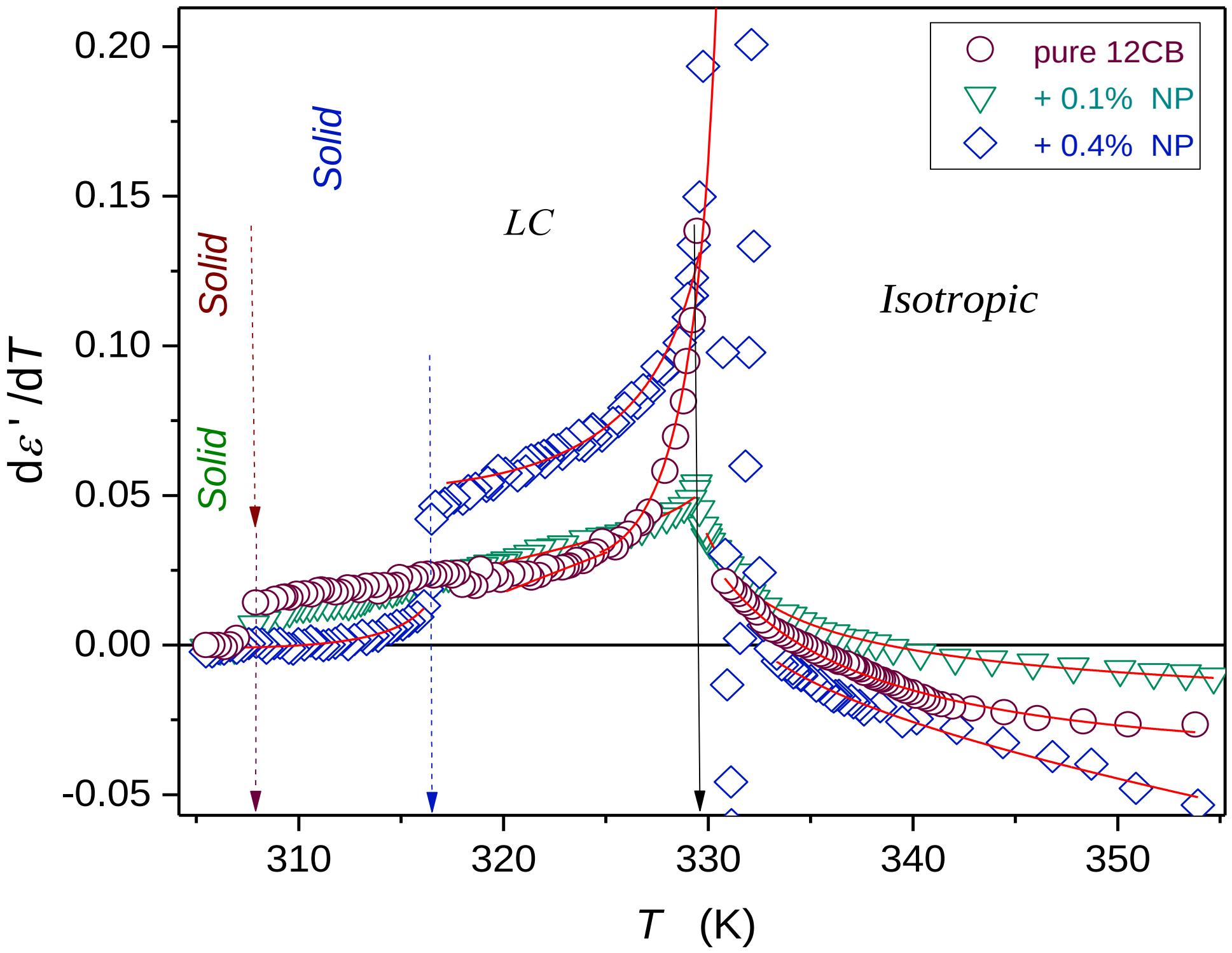

**FIG. 3. (Color online)** Temperature derivative of experimental $\varepsilon(T)$ data for liquid crystalline 12CB and its colloids with nanoparticles $BaTiO_3$, given in Fig. 1. Red curves show Eqs. (4) and (5). Dotted arrows indicate the isotropic mesophase transitions (clearing temperature). The solid arrow is for the mesophase-solid transition.

The distortion-sensitive nature of the derivative analysis makes it possible to reveal even subtle features hidden within experimental data. For the case tested in the given research such an analysis enables a precise estimation of the range of pretransitional behavior, dominated by multimolecular pretransitional fluctuations. Particularly notable is the clear emergence of the pretransitional effect for the solid to mesophase transition (for $x = 0.4\%$), which is absent in pure 12 CB. It is shown by Eq. 3, with the power exponent $\alpha \approx 0.5$ and the small value of discontinuity $\Delta T \approx 3$ K. For all concentrations of nanoparticles, the clearing temperatures are indicated by dashed lines with arrows, indicating the jump of the dielectric constant associated with the subsequent isotropic-mesophase and mesophase-solid transitions.

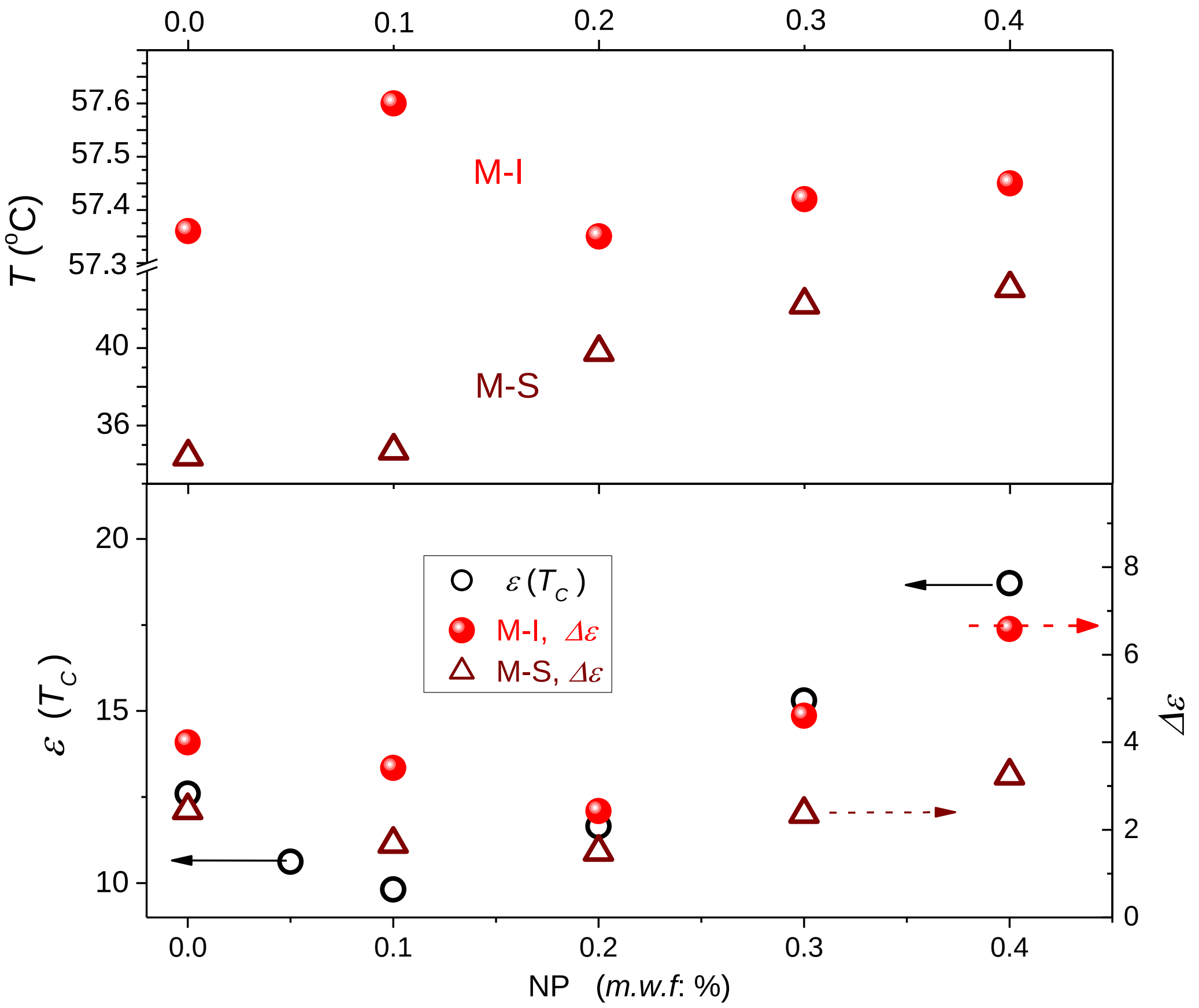

**FIG. 4. (Color online)** Concentration dependence of $BaTiO_3$ nanoparticles [mass weight fraction MWF×100%] of the dielectric constant in the isotropic phase at the clearing temperature $\varepsilon(T_C)$ (open circles, left scale) and jumps of the dielectric constants $\Delta\varepsilon$ (closed circles, right scale) at the SmA-I (closed circles, right scale) and SmA-S

(triangles, right scale) phase transitions. The *M* is for the LC mesophase, *I* is for the isotropic liquid and *S* is for the solid phase.

Figure 4 summarizes the impact of NP concentration on 12CB. The upper part shows the influence on the isotropic – mesophase and mesophase – solid phase transition temperatures. Notable is the almost negligible impact on the first one and the very strong impact on the second one. The lower part of Fig. 4 shows the dependence on concentrations of the dielectric constant in the isotropic liquid phase (for $T = T_C$) and for the change (jump) of dielectric permittivity for subsequent phase transitions. The values of these parameters initially decrease greatly (by approximately 50% in comparison with pure 12CB) and then greatly increase (by approximately 50% in comparison with pure 12CB).

## IV. CONCLUSION

This Rapid Communication presented an analysis of pretransitional anomalies in LC compound + nanoparticles colloids, focusing on such basic characteristics as the power (critical) exponent, the discontinuity of the phase transition and temperatures of subsequent phase transitions. Studies were carried out in 12CB with isotropic→Sm-A→solid mesomorphism, based on dielectric constant measurements. It is notable that obtaining stable nanocolloids does not required additional agents in the tested range of NPs concentrations. Dielectric constant studies made it possible to monitor the tendency of changing the average arrangement of a permanent dipole moment coupled to rod-like molecules of 12CB from the parallel ($d\varepsilon/dT < 0$) to the antiparallel ($d\varepsilon/dT > 0$) one. In the isotropic liquid this is associated with the appearance of premesomorphic fluctuations linked to orientational ordering of molecules associated with the antiparallel arrangement of the permanent dipole moment to minimize the energy of formation of pretransitional fluctuations. The growing size and number of fluctuations causes the number of molecules ordered in the antiparallel way to increase over the chaotic *fluidlike* surrounding. Consequently, $\varepsilon(T)$ decreases on approaching the clearing temperature and the mesophase – solid phase transition.

In LC mesophases the orientational ordering with antiparallel ordering of the permanent dipole moment is the background feature. However, on approaching the weakly discontinuous ($M \to I$) phase transition one can expect the appearance of *prefluidlike* fluctuations with chaotically oriented molecules within the mesophase. Consequently, the average dielectric constant of the system should increase, which indeed takes place in Figs. 1 and 3. The same mechanism can be expected for the solid to mesophase transition. However, in the solid crystalline state, an almost perfect orientational antiparallel arrangement of permanent dipole moments can be expected. The appearance of premesomorphic fluctuations with much less order in the antiparallel way of rod-like molecules has to lead to the pretransitional increase of dielectric constant. The opposite phenomenon should be expected for the mesophase to solid transition, leading to the pretransitional decrease of the dielectric constant. Both phenomena are visible in Fig. 3. However, it is notable that they clearly emerge only when introducing the frustration factor associated with $BaTiO_3$ nanoparticles. Their presence also shift the isotropic to mesophase transition towards the notably less discontinuous case, reducing notable values of discontinuities $\Delta T^*$ and $\Delta T^{**}$. Contrary to the existing evidence (mainly for the I-N transition) no influence of nanoparticles on the clearing temperature was detected. However, with increasing concentration of nanoparticles, a strong increase of the solid to mesophase transition temperature takes place.

The addition of nanoparticles also very strongly influences the average value of the dielectric constant: (i) For a small concentration of NPs ($x = 0.1\%$) it decreases by approximately 50 % and (ii) for increasing concentration up to $x = 0.4\%$ it increases by approximately 50 %, in comparison with pure 12CB. This suggests that for smaller concentration of $BaTiO_3$ NP interactions with large permanent dipole moments of 12CB can facilitate some background orientational and antiparallel arrangement of 12CB molecules. Consequently, the average value of the dielectric constant decreases. A notably increase of NP concentration distorts the antiparallel background arrangement, leading to an increase of the very average value of the dielectric constant (x = 0.4%)

For each phase (the isotropic liquid phase, mesophases and the solid state) the same value of the critical exponent was obtained. This suggests a lack of the impact of the nanoparticles on universal properties of phase transitions in the discussed case. It is also notable that the description related to the pretransitional behavior extends to the whole temperature range tested in the isotropic liquid, mesophase and solid state. This indicates that at least for the dielectric constant, the behavior in all phases is dominated by collective multimolecular pretransitional fluctuations. Concluding, the results of this Rapid Communication show that adding nanoparticles can present opportunities to reach deeper insight into the fundamental properties of phase transitions, particularly in LC materials, with possible impact on practical implementations.

**ACKNOWLEDGMENTS**

This study was supported by the National Science Centre (Cracow, Poland) via Grant No. 2011/03/B/ST3/02352 and No. 2011/01/B/NZ9/02537.